\documentstyle[12pt]{article}
\newcommand\beq{\begin{equation}}
\newcommand\eeq{\end{equation}}
\def\beqa{\begin{eqnarray}}
\def\eeqa{\end{eqnarray}}
\def\bega{\begin{array}}
\def\enda{\end{array}}

\def\non{\nonumber}

\advance\oddsidemargin -0.5cm
\advance\textwidth 1cm
\begin{document}
\title{Clocks and Time}

\author{Arlen Anderson\thanks{arley@physics.unc.edu}\\
Dept. Physics\\ UNC-Chapel Hill\\  Chapel Hill NC 27599-3255}
\date{July 17, 1995}

\maketitle
\vspace{-8cm}
\hfill IFP-UNC-95-512

\hfill gr-qc/9507039
\vspace{7cm}

\begin{abstract}
A general definition of a clock is proposed, and
the role of clocks in establishing temporal pre-conditions
in quantum mechanical questions is critically discussed.
The different status of clocks as used by theorists external
to a system and as used by participant-observers within
a system is emphasized.  It is shown that the foliation of
spacetime into instants of time is necessary to correctly
interpret the readings of clocks and that clocks are thus insufficient
to reconstruct time in the absence of such a foliation.
\end{abstract}
\newpage

``How does one know what time it is?''  The standard retort is,
``by consulting a clock!'' This answer is probably as old as the quantum
theory itself, but is that answer glib, or does it withstand
scrutiny?  What is a clock, and does it faithfully reflect time?

This question, how one knows what time it is, is a central challenge
in quantum gravity and underlies many of the problems of time and
the proposals to resolve them\cite{Kuc1,Ish1,Ish2}.
Recently Kucha\v{r}\cite{Kuc2} has criticized Rovelli's
``evolving constants of motion''\cite{Rov1} with a question closely
related to this, and this is discussed in \cite{Andother}.
The prominent need to identify time arising in Wheeler's
program\cite{Whe} of the ``3-geometry as the carrier of time''
is well known, but a discussion focussing on the Wheeler-DeWitt equation is
subject to unnecessary technical distraction.  On the other hand, virtually
all questions in quantum mechanics implicitly carry time preconditions, so
the need to ``know'' time is ubiquitous.
In view of the broad significance of this question and recent interest
in clocks (e.g. \cite{Mar}), a general discussion of the relation between
clocks and time is called for.

\section{The Nature of Clocks}

When one asks, ``what is the position of particle-a?'', one implicitly
assumes ``at time $t$.'' But how does one know what time it is?
By consulting a clock?
Consider the question, ``what time does the clock show?'' Necessarily the
question again implicitly assumes ``at time $t$,'' but one cannot now
answer that one knows the time $t$ by consulting the clock. Instead
one relies on a primitive immediacy\cite{now} in the understanding that it
is ``at this moment as I ask the question,'' which becomes ``at this moment
as I make the observation'' when one seeks the answer.

When talking about time in quantum physics, it is prudent to distinguish
two points of view about time which are wont to be conflated. One is the
time of the theorist who has access to the full wavefunction of the
universe. The other is the time of a participant-observer who must make
measurement-observations of other subsystems to obtain markers of
particular instants. It is the participant-observer who must rely on
primitive immediacy, or the sense of ``now.'' The theorist can look at a
map of the spacetime history of a system and, pointing with a finger, ask,
``what time does the clock show---here?'' Neither can answer in any
absolute sense what time $t$ it is, not at that moment ``now'' nor at that
place in spacetime, but then they don't need to. Time itself is not being
measured because no value for $t$ is returned by a measurement-interaction.
One does not need to know $t$ to be able to measure an
observable: one can measure the self-adjoint operators which are convenient
to measure, and these are necessarily instances of
observables\cite{Andother}.

One of the challenges of dealing with time in quantum cosmology is to
reconcile these two points of view.  As theorists, our goal is to find
the map which is the spacetime history of the wavefunction of the
universe.  And, as participant-observers within the universe, our desire
is to translate that map into an understanding of what we may
experience.  We must be able to move back and forth between the two
points of view comfortably.

There are cautionary difficulties however. We are far more limited in our
observational capacity as participant-observers than we are as theorists.
Given the full wavefunction of a system, it is easy to see the details of
the states of its component subsystems. As one of those subsystems, it is
far more difficult to deduce that information through
measurement-observations. In particular, there are limitations on the
nature of couplings that we are allowed as observing subsystems.
Furthermore, it is in the nature of theorists to idealize component
subsystems to their ``essential'' elements and in particular to minimize
their numbers of degrees of freedom. These features of the different realms
run at cross purposes.

For instance, there is a danger of confusing the difficulty of the
participant-observer to learn some piece of information that is obvious
to the theorist with the absence of that information.  In the
participatory realm, one must remember that not all observers are
physicists and they have no need to ``know'' more than it takes them to
evolve properly under the dynamics.  Thus, suppose one has a system
consisting of a clock-subsystem and subsystems -a and -b.  Subsystem-a does not
have to have the necessary couplings to be able to measure the clock's
state in order to interact with subsystem-b nor for time to pass.  When
the theorist speaks of subsystem-a as the observer, and says that it
makes a measurement when the clock reads, say $2{:}00$, subsystem-a does
not have to verify (or be able to verify) the truth of that statement
when it makes the measurement.

When discussing time, it has become popular to insert clocks
into systems to label time or to speak of certain subsystems as being ``the
clock.''  This is an unnecessary device of convenience
to aid the theorist.  Time does not become any more real because there
are readily identifiable clocks present in a system.  This will become
clearer when clocks are defined and discussed below.

Turning the situation around, the theorist must be wary of the fact that
models with few degrees of freedom may impose limitations that are not
present in larger systems.  For instance, suppose that the theorist is
concerned that the observer-subsystem be able to read the state of a
clock many times to observe the passage of time.  If the full system is
not sufficiently complicated, the theorist may find that the interaction
between the observer and the clock ``saturates'' so that no further
distinct measurements are possible.  This is an artifact of the idealized
nature of the system.  It should be a concern whenever one insists on
modelling all interactions.

Following these general remarks, it should not be disturbing that
there are theorems that time cannot be observed by measuring a
self-adjoint operator\cite{UnW,Pau,Har}. More
precisely, there is no self-adjoint operator whose expectation value increases
monotonically in correlation with time for quantum systems
whose Hamiltonian is bounded from below. Generally one finds that there is
a non-vanishing amplitude that the putative time will run backwards.
Since we have no great direct need as participant-observers to measure time
itself, our inability to do so is not an insurmountable loss.

Our desire to measure time is driven by our perception of dynamics as
theorists. We think of dynamics as parametrized by a time $t$, and we want
to coordinate this time with the flow of (participatory) physical time in
order to translate our theoretical considerations into practical
predictions. Our access to time even as theorists is more limited than we
may be accustomed to believe. This is because our experience with Newtonian
dynamics creates an intuition about a rigidity to time and spacetime
slicing which is not present in a time-reparametrization and diffeomorphism
invariant theory like general relativity. Of course we recognize that an
easy way to deal with this is to look for diffeomorphism invariant markers,
that is, for correlations amongst physical states. This is how clocks are
often used: they are time-markers. One can identify where one
is in the time continuum by specifying one's position in relation to a
sequence of clock readings.

In the context of quantum theory,  one does not even need
to ``know'' one's relation to the time-markers or clock states; it appears
sufficient for there to be a condition of correlation. The idea here\cite{PaW}
is that much of quantum mechanics is based on conditional statements like
``if spin-b is down, then spin-a is up with certainty.'' Such statements
are the post-Everett interpretation of the meaning of product states like
$\left|\uparrow\right\rangle_a \left|\downarrow\right\rangle_b$. If our
knowledge of time is simply through condition of correlation, then
statements involving time reduce to
familiar conditional statements of quantum mechanics: ``if the clock reads
2 o'clock, the particle is at position $x$.''

In the conditional probability interpretation, one does not measure time by
computing the expectation value of a self-adjoint operator against the
state of the clock (as assumed in the no-go theorems). Rather one
understands in the post-Everett sense that such measurement has implicitly
taken place when ``a correlation is established through interaction.'' The
conditional statement ``if the clock reads 2 o'clock, the particle is at
$x$ with certainty'' is the interpretation of the product state
$|x\rangle|2{:}00\rangle$.
Admittedly, the status of conditional statements in quantum mechanics is
not universally agreed upon, with a large segment of the community holding
that a major part of the measurement problem in quantum mechanics is to
explain how one establishes the truth of pre-conditions in conditional
statements. But, at least one hopes that statements referring to time are
no worse than any other conditional statements. This is what we will
investigate.

Let us examine this picture of conditional correlation more closely. In the
correlation of a pair of spin one-half states, one can pose and answer the
linguistically symmetric questions ``what is the state of the first spin
given that the second is down?'' and ``what is the state of the second spin
given that the first is down?'' These questions are on an equal footing
physically. Contrast this with the questions ``where is the particle at
time $t$?'' and ``when is the particle at position $x$?'' These questions are
linguistically symmetric, but they are not on an equal footing physically.
The former question can be answered by giving a probability distribution,
but it isn't clear what answer if any can be given to the second in
conventional quantum mechanics.

In quantum mechanics, it is the existence of exhaustive and exclusive sets
of alternatives that allows us to give probability distributions as answers
to questions, or more precisely as predictions of the outcomes of
experiments to answer questions. With spin one-half states, when the spin
is measured in the z-direction, it can either be up or down but not
both. In each measurement, some alternative must occur, but no
more than one does occur. For a particle position at a given time, the
following statements are satisfied
\begin{itemize}
\label{seea}
\item Every object is somewhere.

\item No object is in more than one place.
\end{itemize}
The first statement assures that position at a moment of time is
exhaustive and the second that it is exclusive.  The analogous statements
about position in time at a given spatial location are not generally
satisfied.  An object need never be at a particular place or it
may be there many times.

That alternatives in temporal location at a given place
do not form an exhaustive and exclusive set indicates a fundamental
asymmetry between space and time in conventional quantum mechanics.
It is easy to trace this asymmetry to the fact that states are
essentially superpositions of alternatives at moments of time
(or on spacelike hypersurfaces, which correspond to the notion of a
moment of time in curved spacetime).  This reveals something about
the correlation picture of clocks.  Since conditional
statements about a pair of physical states are symmetric, statements about
a system and its clock are symmetric.  This is because the
possible states of the clock and the states of the system both come
from exhaustive and exclusive sets.
The implication is again that a clock is not a direct reflection of time.

As a precise and general definition of a clock, I propose the following: a
clock is any subsystem with whose motion another subsystem may be
correlated.
In what sense does this definition describe a clock? Perhaps the most
fundamental feature of time\cite{Rov2} is the uniqueness and
distinguishability of different moments of time. Unruh and Wald\cite{UnW}
refer to this as the Heraclitian nature of time. In product states
representing the correlation of one subsystem with another, in so far as
the motion of one subsystem passes through distinct states, these reflect
distinguishable moments of time. The subsystem acts as a clock in that it
allows moments of time to be distinguished.

By this definition, almost everything is a clock. There is however
a difference between how a clock is viewed by a theorist and
by a participant-observer.  A glass of water standing on a table serves as a
clock because its internal quantum state varies from moment to moment.  From
my perspective, looking at it with my eyes, I cannot perceive the
differences between these states.  I can only appreciate the glass of
water as a clock on the much longer time scale when there are changes of
water level through evaporation.
This leads to the conclusion: to a theorist, anything can serve as a
clock, but an observer won't necessarily be able to read it!

Interestingly, in conflict with the desire to distinguish instants,
we are accustomed to associate
``good'' clocks with periodicity, in which a given physical state of a
clock refers to a sequence of times. I would argue that periodicity is
important for a secondary function of clocks, namely for measuring uniform
intervals of time. Uniform periodicity is useful in synchronizing and in
verifying accuracy of clocks. It is by observing other factors that
periodicity is broken and distinct moments are recognized: the minute hand
advances after each cycle of the second hand; the date changes each time
midnight comes again. If one were not concerned with measuring intervals of
time, but only with distinguishing moments, periodicity would not be
important and indeed might be a negative feature.

Another attribute widely expected of clocks is the ordering of the instants
of time. This is not implied by the definition of clock given
here. In so far as a subsystem can be seen as moving through a sequence of
states under deterministic
evolution, this motion provides an ordering to that sequence of states and
hence to the moments of time they label. Independent of this motion
however, there is not generally an ordering on the possible states of the clock
subsystem. This is important for example in (classical) general relativity
where each spacelike slice through a 4-manifold solution of the Einstein
field equations is an instant of time, the clock-state being the 3-metric
and its conjugate momentum on that slice. There is at best a partial
ordering of these spacelike slices. One can choose to focus on foliations
which have a definite ordering, but clock-states are defined by every
spacelike slice.

Returning to the issue of periodicity, consider the correlation question,
``where is the particle when the clock
reads 2 o'clock?'' The answer to this question is a member of a set of
exhaustive and exclusive alternatives only if one implicitly assumes that
one means one of the unique moments when the clock reads two o'clock. If
the unique moment is not identified, the particle may be in more than one
place. The problem is the same as that above when asking ``when is the
particle at position $x$?'' This is important because it undercuts the
conditional correlation picture. The conditional is only meaningful when
identification of the unique moment is assured. Then one is guaranteed that
there are an exhaustive and exclusive set of alternatives.


Ultimately, the uniqueness of each moment of
time is reflected only in the uniqueness of the state of the full system,
including all its clocks. Because the equations of motion for states are
deterministic, if the initial data for the full system reoccur, the
exact evolution of the system repeats, and in such case time is truly
periodic.
Suppose however that attention is restricted to a subsystem, as it
usually is in practice.
It is always possible to arrange, by adding appropriate
degrees of freedom external to the subsystem and controlling
their interaction with the subsystem, that the
subsystem repeats its state without the state of the full system
repeating. Time itself does not repeat, but one cannot use the
subsystem to confirm this.  Thus, no subsystem in
isolation can be wholly reliable as an indicator of the passage of time.

By introducing the notion of clocks, one has made the measurement of time
and space seem more symmetric, but the apparent progress only reveals
further trouble. The measurement situation is indeed symmetric, but where
before there once was a problem with conditionals based at a position in space,
now there is a problem with conditionals specified at a moment of time
because subsystem clocks do not always successfully identify unique moments.
The original
question stands only mildly modified: how does one identify the unique
moment of time? I have given the answer above: the unique moment of time is
only unambiguously identified by the full state of the system. This is not
something one can set out to measure from within. There is no Great Clock
Variable that is present in all situations which we as
participant-observers can consult to see what
time it is. We are forced to accept that our ability to measure time is not
absolute.

Is this really a problem?  No.  As participant-observers,
we have a perception of unique
moments and we use that perception to improve on our clocks.  Truly we
may only be consulting a biomechanical system with a large number of
degrees of freedom which is not locally periodic, but that stands
adequately as a representative of the full state of the universe.  That it is
possible in principle for any subsystem to repeat its state
without the full system doing so, and that therefore no subsystem can be
assured of being a perfect clock, is of no concern---if we are not
looking for a Great Clock Variable! We are working
within a small spacetime domain and we only require that our measurement
of time is adequate within that domain.

This argues that we should be satisfied with our clocks even though we can
construct mathematical examples in which models of them are inadequate. The
impudent answer by a participant-observer to the question of how one
knows what time it is, is that ``I {\it know} what time it is when
I consult a clock.'' I know this in the sense that I have the perception
that I can distinguish that moment from other moments. A theory of
consciousness is not needed, only a non-periodic subsystem with
sufficiently large numbers of degrees of freedom to distinguish
between the different periodic moments of the clock. Does
something have to explicitly keep track for the moments to be different?
No, it is sufficient that the state of the full system is different. If you
like, the state of the environment is what guarantees that clocks refer to
unique moments. Our internal perceptions are simply a symptom that the
environment of the clock is different each time it repeats its state.

One might even be tempted to say that the environment decoheres the different
instants of time, or more accurately that it decoheres the different clock
readings so that they distinguish different moments of time.
This is an intriguing picture, but one must be careful if one tries to take it
literally.  Different clock states occur at different moments of time
and while one may compute matrix elements between them, this is not what
one would do in the usual (Zurek) density matrix decoherence picture.


By inspecting the evolution of a particular full system, one may determine
that the state of some subsystem does not repeat. That subsystem then
serves as a clock to distinguish instants for the full evolution,
but it is only known to be adequate because the full evolution has been
consulted. The point here is that there is no single Great Clock Variable
that will serve as a clock in all systems. It is futile then to look for
such a time variable, be it an intrinsic or extrinsic time, to interpret
the wavefunction of the universe obtained from the Wheeler-DeWitt equation.
I emphasize that this is not to say that time does not exist on super-phase
space (the space of 3-metrics and their conjugate momenta). It
does, but it
is not so easily captured as to admit representation in a single functional
combination of the 3-metric and its conjugate momentum for all
spacetimes.  It seems entirely plausible given the freedom in initial data
that any such functional combination can be made to repeat by suitable
choice of initial data.

\section{A {\it Gedanken} Experiment}

To illustrate the distinction between a theorist's clock and one of
a participant-observer, consider a free particle-a and a subsystem-b,
described by the Hamiltonian $H_b$, which do not interact. The
super-Hamiltonian is
\beq
{\cal H}= p_0+{1\over 2}p^2 +H_b.
\eeq
Since particle-a is free and uncoupled from subsystem-b, it can be prepared
at $t=0$ as a normalized Gaussian, $\psi_a(q,0)=\pi^{-1/4}\exp(-q^2/2)$.
The variance of this Gaussian will grow as the state evolves.  The full
system can be prepared as a product of this state and the evolving state
of subsystem-b
\beq
\Psi(q,x,t)=\psi_a(q,t)\phi_b(x,t).
\eeq
The point is that, by the definition above, particle-a acts as a clock
because its state distinguishes moments of time. It is however a clock
which is never physically read by subsystem-b. It is a theorist's clock.
One can use it to pose conditional questions: ``what is the state of
subsystem-b when the clock-state (particle-a) is $\psi_a(q,t_0)$?'' Note
that it is a perfect clock even though it is described by a Hamiltonian
which is bounded from below. This does not violate the theorems quoted
above because no self-adjoint operator is introduced to attempt to read the
time by taking expectation values.

This clock is used differently than one might naively expect. One doesn't
compute the conditional by projecting $\psi_a(q,t_0)$ on
$\Psi(q,x,t)$---there would be nonzero overlap for all $t$. In a very real
sense, the sequence of states $\psi_a(q,t)$ are the moments of time. They
are the only readings the clock is allowed to take and they uniquely
correspond to temporal locations in the evolution of the full system.

Consider the situation when subsystem-b is not in a product state
with the clock.  Take the simple superposition
\beq
\Psi'(q,x,t)=\psi_a(q,t)\phi_b(x,t) +\psi_a(q,t+\Delta)\phi'_b(x,t).
\eeq
The clock-state associated to the second term in the superposition is
the same as in the first term only advanced by time $\Delta$.  Here, the full
system state denotes a unique moment of time, but the clock subsystem
seems to refer to more than one instant.  Suppose one asks what the state
of subsystem-b is when the clock-state is $\psi_a(q,t_0)$.  This question
only makes sense if it is completed by the information of what time $t$
it is!  If the time were $t=t_0$, the state would be $\phi_b(x,t_0)$.  If
the time were $t=t_0-\Delta$, the state would be
$\phi'_b(x,t_0-\Delta)$.  One needs sufficient additional information to
determine which eventuality occurs.  This hints at an essential
inadequacy of clocks as surrogates for time.

To amplify on this, consider a {\it gedanken} experiment in which this
situation could arise. Let an observer-scientist experience the experiment
as part of a subsystem-a. Suppose that subsystem-a (the scientist and her
equipment) is initially coupled to both subsystem-b and clock-c so that it
can monitor both. For definiteness, let clock-c be an ordinary (nearly
classical) mechanical clock, and take subsystem-b to be a spin one-half
particle precessing in a magnetic field. Assign the rest of the universe
to subsystem-e (the ``environment'').

As an element of subsystem-a, the scientist has a sense of ``Now.''
Occasionally she looks at clock-c, and she observes a sequence of readings
with which she chooses to label the sequence of ``Now'' instants. She also
inspects subsystem-b and prepares it in an initial state. At this moment,
the initial state of the full system is a product state of the states of
clock-c and subsystems-a, -b and -e.

The scientist puts subsystem-b and clock-c in a rocket and then closes
herself in her lab-bunker---which does not contain a clock---preparatory to
doing a Schrodinger-cat-like experiment. Her experiment consists of a
trigger (consider it to be part of subsystem-e) which takes the form of a
spin one-half particle with spin up in the $z$-direction entering a
Stern-Gerlach apparatus with magnetic field in the $x$-direction. If the
spin leaves to the right (spin up in the $x$-direction), nothing happens.
If it leaves to the left, clock-c and subsystem-b are launched in the
rocket and later return to the lab. The rocket completes its journey
sufficiently fast for its clock to be $1$ minute behind what a clock
would read that did not make the journey (i.e. a twin-paradox effect).

The scientist waits a sufficiently long, but unknown, time, and leaves her
lab. Prior to opening the rocket, the state of the full system is
\beq
2^{-1/2}|S;t\rangle_a (|\omega t\rangle_b |t\rangle_c |+x,{\cal E}\rangle_e
+ |\omega(t-1)\rangle_b |t-1\rangle_c |-x,{\cal E}'\rangle_e).
\eeq
When the scientist opens the rocket and looks at clock-c, she predicts
the precession of the spin in subsystem-b. She then verifies that she is
correct. Even though she can make this conditional probability calculation
correctly, she cannot determine whether the rocket took the journey or not
without looking at subsystem-e. Therefore she has no way to determine how
much laboratory time has elapsed. As one expects from special relativity,
clock-time is not an absolute measure of time.  Clock-c has a
known correlation with the motion of subsystem-b and is therefore sufficient
to determine that the time-precondition of a quantum mechanical question is
satisfied.  Clock-c does not have a known correlation with subsystem-a,
the laboratory, and cannot be used to establish time-preconditions.

Note that the scientist's sense of ``Now'' is also not a measure of
time---it distinguishes instants, but it does not measure intervals.
The measure of time is one of the attributes of clocks beyond
merely distinguishing instants.

Consider a second scenario. The scientist leaves her bunker, inspects
sub\-system-e and determines that the rocket took its journey. Then knowing
the path of that journey, she computes the twin-paradox effect. When she
looks at the clock, she can deduce how much laboratory time has passed.
Clock-c becomes sufficient as a measure of laboratory time because she
knows how its motion correlates to the motion of the laboratory.
One concludes that in order to use a clock to establish
the time pre-condition of a quantum mechanical
question it is necessary that the motion of the clock have a
known correlation with the subsystem of interest.

As the final and most revealing scenario, suppose that during the rocket's
journey, it flies
through a magnetic field which perturbs the spin-precession experiment. The
scientist does not know about this because it is an unmodelled consequence
of interaction between subsystem-b and subsystem-e. For definiteness,
suppose that at 2{:}00 in the rocket which did not make the journey, the
spin is up in the z-direction while at $2{:}00$ in the rocket which did make
the journey, it is down.  Because of the twin paradox delay, there are
two spacetime events involved, say $t_1$ and $t_2$ in terms of an
external labelling by a theorist.
The state of the clock, spin and scientist at these times are
\beqa
\label{twotimes}
t_1&:&\quad 2^{-1/2}(|S_1;t_1\rangle_a |\uparrow\rangle_b |2{:}00\rangle_c
|{\cal E}_1\rangle_e +
|S'_1;t_1\rangle_a |\downarrow-\epsilon\rangle_b|1{:}59\rangle_c
|{\cal E}'_1\rangle_e) \\
t_2&:&\quad 2^{-1/2}(|S_2;t_2\rangle_a |\uparrow+\epsilon\rangle_b
|2{:}01\rangle_c  |{\cal E}_2\rangle_e
+ |S'_2;t_2\rangle_a |\downarrow\rangle_b |2{:}00\rangle_c
|{\cal E}'_2\rangle_e) \non
\eeqa
As outsiders who are privy to the wavefunction of the full system and to an
external time labelling, we can pose and answer the question, what is the
probability, after the scientist looks at the clock, that she will
correctly predict the spin-precession? The answer of course is that it
equals the probability that the rocket did not take the journey.

Suppose she looks at clock-c, but not at subsystem-b. We pose the
conditional question, ``what is the state of subsystem-b when she sees that
clock-c reads 2{:}00?'' This is the same peculiar situation as above where
there was a superposition state containing a free particle-clock at two
readings. She will see clock-c read 2{:}00 at two different spacetime
locations, depending on whether the rocket made its journey or not. This is
not a conditional question posed at a single instant of time. The clock
does not act effectively to specify a moment of time.

I emphasize that this has happened even though the clock is, for all
intents,
perfect. The clock is effectively classical, and nothing disturbs it during
the course of the experiment. Any trouble cannot be attributed to the clock
itself. Clearly, a similar outcome could be achieved with a defective clock
or a clock whose evolution is disturbed by some quantum evolution, but then
one might argue that the situation reflects some failure in the clock.
Here, there is no such failing and yet the clock does not fulfill its
function from the perspective of one outside the system.

Certainly, within the system, the scientist can perform her experiment of
reading the clock and checking the spin, and she will find definite
outcomes. She will be surprised that the clock is not correlated with the
spin in the branch where the rocket made its journey, but that is the
extent of it. If she repeats the experiment many times, she will discover
that the correlation of clock and spin is broken half of the time. By measuring
other directions of the spin, she can determine that the state of the spin
is incoherent. As a participant in the time flow of the universe who
therefore has a sense of the passage of time, nothing especially unusual
about this situation is apparent to her. It is in the attempt to inspect
the full system from outside that there are potential difficulties.

\section{The Inadequacy of Clocks}

To comprehend the significance of this situation, consider the situation as
analogous to what one confronts in a traditional Wheeler-DeWitt approach to
canonical quantum gravity. One is given a super-Hamiltonian constraint.
This argument is not really about gravity, so the super-momentum
constraints can be ignored. For definiteness, suppose the super-Hamiltonian
describes an experiment similar to the {\it gedanken} experiment above.
There is a clock correlated to a spin, and there is a quantum event which
produces a motion leaving the clock-spin subsystem in a superposition
with the same clock reading occuring at different
spacetime locations. Suppose that one succeeds in solving the
super-Hamiltonian constraint corresponding to the described experiment
to obtain the wavefunction of the universe. The difficulty one
faces is that one does not know which variable or combination of
variables in configuration space is time. More particularly, one doesn't
know what constitutes an instant of time, so one doesn't know how to
foliate spacetime into spacelike slices. This means that one doesn't know
how to classify alternatives into exhaustive and exclusive sets.

One turns to the notion of a clock. The proposal has been made to inspect the
state of the full system and identify the state of subsystem-c, the clock.
Because of the dynamics of the system, this state is correlated to the
state of subsystem-b and seems a likely surrogate for time. Inspecting the
wavefunction of the universe, one sees that there are two events where
clock-c reads 2{:}00 and that at these events the state of the spin is up
in one and down in the other. One wants to know what this says about the
state of the spin when the clock reads 2{:}00. Is the state
``$2^{-1/2}(|\uparrow\rangle_b+|\downarrow\rangle_b)$'' or is it
``$|\uparrow\rangle_b$ or $|\downarrow\rangle_b$ with probability 50\%''?
One cannot answer this unless one knows whether the events labeled by
2{:}00 occur on the same spacelike hypersurface or on different
hypersurfaces. But one doesn't know this, and the clock can't help!

One is forced to conclude that clocks cannot stand in place of the
knowledge of the spacelike foliation on which exhaustive and exclusive
alternatives are defined.  A clock can have the same reading at more than
one spacetime event.  If the events are spacelike separated, one expects
a coherent superposition of alternatives, given a clock reading.  If the
events are timelike separated, one expects an incoherent collection of
alternatives.  It may even be that when there are an infinite number of
timelike separated events with the same clock reading, one cannot
normalize a probability distribution for the outcomes correlated to that
reading.

A clock is a useful device for distinguishing moments of time for
participant-observers who have an ``experiential'' sense of time. This is in
the sense that by virtue of distinctions in their own state, they can
distinguish moments, even if they are not ``consciously'' self-aware. The
usefulness of clocks in interpreting the wavefunction of the universe is
more limited. In particular, it relies crucially on knowledge of the
Hilbert space structure and the associated foliation of spacetime into
instants of time. Knowledge of the instants of time is necessary to collect
alternatives in exhaustive and exclusive sets. It is also necessary to
determine whether the correlation with a given clock reading is coherent or
not.  Clocks cannot be used to deduce the Hilbert space structure or the
foliation of spacetime into instants of time.

\section{Quantum Spacetime Events}

An alternative worth considering is whether conventional quantum mechanics
can be modified in such a way that conditional statements involving time
always have well-defined answers. This would restore the symmetry between
space and time. This could be accomplished if there were exhaustive and
exclusive sets of alternatives that refer to time. It was mentioned above
that the temporal analog of the statements about positions in space are
untrue. If one
deals with events in spacetime as the analog of positions in space, then
one can formulate an exhaustive and complete set of alternatives. One has
the statements
\begin{itemize}
\label{eveea}
\item Every event is somewhere in spacetime.

\item No event is in more than one location in spacetime.
\end{itemize}
By virtue of these, one can assign a probability distribution to the
location of events in a spacetime manifold.

Comparison with the spatial analog shows that working with these is
equivalent to quantizing the location of the spacetime events in a
manifold. Because every event must be somewhere and can't be in more than
one place, one can assign an amplitude to the location of events. To
formulate a quantum theory of this amplitude, one can associate four
spacetime ``coordinate'' variables and their conjugate momenta with each
event. These can be taken to satisfy canonical commutation relations (at
least in the simplest case). The eigenvalues of the coordinate variables
would refer to the location of their event in spacetime. This is in analogy
to the spatial case where three spatial coordinate variables refer to the
spatial location of each particle.

A simple model illustrating this is a discrete lattice model of
spacetime.  Each point of the lattice is an event.  Because each event is
quantized, there is a wavefunction describing the displacement of the
actual event in the manifold from its lattice site.  The wavefunction of
the full lattice is the product of the wavefunctions at each site.  This
gives one a truly quantum spacetime geometry.  One can pose questions
asking about the location of a given event and there will be a {\it
bona fide} probability distribution as an answer.  The details of the
wavefunction of the lattice (``wavefunction of the universe'') will of
course depend on the equation which relates the wavefunctions at
different sites.

It is not difficult in this picture to imagine how time
slices and quasiclassical geometry arise when the fluctuations of events
around their lattice sites is small.  It is more difficult to imagine how
matter degrees of freedom could be put into this dynamical background and
be made to evolve.  While quantizing the location of events makes the
interpretation of the quantum spacetime picture tractable from the
outside, one is faced with the issue of how it would appear
from within the system.  One does not expect that there are
operators which measure the location of particular spacetime events.  To
other subsystems, then, spacetime events should appear as identical particles,
possibly fermions so that no two events can occupy the same location in
the manifold.  A correlation picture seems natural in which matter states
are correlated with states for collections (regions?) of events.  This is
admittedly vague, and
further work is obviously needed to make these ideas precise.

Such a picture could not be elaborated into a realistic toy model of
quantum gravity unless one were able to meet two serious criteria. First
one requires that when quantum fluctuations are turned off and events lie
in the equilibrium positions of some quantum potential, they take positions
such that the lattice corresponds to a solution of Einstein's field
equations. The ability to do this is difficult because of the second
problem: the theory should not depend on an {\it a priori} background. In
particular, one should not measure displacement of an event from its
lattice site in a fixed background, else the final structure will depend on
this background. The requirement of a background independent formulation is
a serious obstacle to parlaying this picture into a model of quantum
gravity, even at the level of Regge calculus.

At the level of a toy model however, the idea of quantizing spacetime
events is attractive because it fits naturally with the goal of being
able to ask questions in quantum theory which involve the time.  A similar
conclusion was reached independently in recent work of Isham and
Linden\cite{IsL}.  They were studying quantum temporal logic in
connection with Gell-Mann and Hartle's consistent histories formalism.
They found that putting a quantum logic structure on propositions about
histories led to the idea of imposing independent commutation relations
on variables at each spacetime event.  While they have not elaborated on
the meaning of such a quantization, it is clearly related to the
situation described here.

\section{Conclusion}

Time is a central feature of our experience as participant-observers in the
universe, yet we do not measure it directly.  We are accustomed to confirming
our sense of the passage of time by observing the changing state of
various dynamical systems, and we recognize these systems as clocks.
Indeed it is this observation of change which constitutes our recognition
of the passage of time, and in a general sense, any subsystem whose
motion we may correlate with is a clock:  by its changing state, we may
distinguish instants of time.

{}From an external vantage as a theorist inspecting the full wave function
of a system, any subsystem whose motion is correlated with that of another
subsystem serves as a clock for that subsystem, whether the capacity
to observe the correlation is present or not.  Generally, non-trivial
correlation indicates some interaction between the subsystems, so there
is a form of observation between them, but correlation can
be built in through the initial state.  A theorist detects the passage
of time by noting the changing collection of correlations amongst
the subsystems of the full system as the slice on which states are defined
is moved through spacetime.

A subtlety arises when the theorist attempts to reconcile his conception
of time with the perception of the participant-observer.  The theorist
requires a knowledge of the foliation of spacetime in order to view
change among subsystem states.  The participant-observer has a sense
of ``now'' but has no direct knowledge of the foliation.  The common
folklore is that the theorist can use clocks to determine a foliation
of spacetime from which to predict possible experiences by the participant-
observer.  This is false because quantum theory allows for superpositions
of clock states, each correlated with different subsystem states.
When one tries to assert temporal preconditions to a quantum question by
saying ``when the clock reads --,'' there is a potential ambiguity
because a clock may have the same reading at different spacetime events.
One needs to know the hypersurface on which to read the clock state to
determine if these different occurences of the precondition contribute
to the same amplitude or to incoherent probabilities.  This cannot be
decided on the basis of the clock state alone.

Fortunately, one is not left without options.  A crucial element that is
missing in the picture of the theorist is the involvement of the
participant-observer.  The theorist must coordinate his foliation with the
experience of the participant-observer.  This may help to resolve
ambiguities, but further investigation is needed.

A more radical alternative is to consider a modification of quantum theory
which treats space and time uniformly.  By quantizing spacetime events,
one is capable as a theorist of posing well-defined questions involving
time---time and space are put on an equal footing.  The situation from
the perspective of the participant-observer is less clear and needs to
be studied.  This approach while clearly speculative seems to offer hope
of a novel direction in quantum geometry which redresses the traditional
unequal treatment of space and time in the quantum theory.

\section*{Acknowledgements}

I would like to thank A. Abrahams, R. Myers and J. York for many helpful
discussions.  This work was supported in part by the National Science
Foundation grant PHY-9413207.


\begin{thebibliography}{99}
\bibitem{now}
This primitive immediacy may not be satisfying to everyone because it
suggests the personal involvement of some observer, but I submit that
ultimately it underlies any discussion of time as it is experienced, be it
by people or by inanimate objects. There is a tendency
to anthropomorphize observers which associates with them a connotation of
awareness. This connotation is wholly undeserved, and unnecessarily
complicates the language of measurement and observation. I shall speak of
observers even in the absence of human beings. An observer is simply a
subsystem whose state we choose to focus on as it interacts with other
subsystems.
\end{thebibliography}
\end{document}